\documentclass[aps,pra,twocolumn,10pt,floatfix,citeautoscript,longbibliography]{revtex4-1}
\usepackage{amssymb}
\usepackage{amsmath}
\usepackage[pdftex]{graphicx}
\usepackage{dcolumn}
\usepackage{bm}
\usepackage{textcomp}
\usepackage{sidecap}

\usepackage[svgnames]{xcolor}
\definecolor{trueblue}{rgb}{0.0, 0.45, 0.81}
\definecolor{crimsonglory}{rgb}{0.75, 0.0, 0.2}
\definecolor{forestgreen}{rgb}{0.13, 0.55, 0.13}

\usepackage{hyperref}
\hypersetup{colorlinks,linkcolor={forestgreen},citecolor={crimsonglory},urlcolor={crimsonglory}}

\usepackage{stackengine}
\stackMath
\usepackage{graphicx}

\begin{document}

\title{Effective magnetic field induced by inhomogeneous Fermi velocity in strained honeycomb structures}

\author{M. Oliva-Leyva$^1$}
\email{mauriceoliva.cu@gmail.com}

\author{J. E. Barrios-Vargas$^2$}
\email{j.e.barrios@gmail.com}

\author{G. Gonzalez de la Cruz$^1$}
\email{bato@fis.cinvestav.mx}

\affiliation{$^1$Departamento de F\'isica, CINVESTAV-IPN, 07360 Ciudad de M\'exico, Mexico.}

\affiliation{$^2$Departamento de F\'isica y Qu\'imica Te\'orica, Facultad de Qu\'imica, UNAM, 04510 Ciudad de M\'exico, M\'exico}

%\date{\today}

\begin{abstract}
In addition to the known pseudomagnetic field, nonuniform strains independently induce a position-dependent Fermi velocity (PDFV) in graphene. Here we demonstrate that, due to the presence of a PDFV, the Dirac fermions on a nonuniform (strained) honeycomb lattice may experiment a sort of magnetic effect, which is linearly proportional to the momentum of the quasiparticle. As a consequence, the quasiparticles have a sublinear dispersion relation. Moreover, we analyze the general consequence of a PDFV on the Klein tunneling of electrons through pseudomagnetic barriers. In particular, we report an anomalous (Klein) tunneling for an electron passing across velocity barriers with magnetic features. Our findings about the effects induced by a PDFV on Dirac fermions in (2D) strained honeycomb lattice could be extended to (3D) Dirac and Weyl semimetals and/or its analogous artificial systems.

\end{abstract}

\maketitle

\section{Introduction}

One of the key features of graphene is its conical electronic band-structure at the so-called Dirac cones. As a consequence, in this two-dimensional conductor, low-energy electrons behave as massless chiral Dirac fermions~\cite{Novoselov2005,Kim2005}. This special electronic behavior has suggested the possibility of observing Klein tunneling~\cite{Katsnelson2006,Kim09,Stander2009}, originally predicted in the context of particle physics.

Another relevant feature of graphene lies in the unusual influence that the elastic lattice deformations have on its electronic properties~\cite{Vozmediano2016,Naumis2017,Zhai2019}. Nonuniform strains in particular constitute a useful tool to implement the concept of strain engineering in graphene because they may induce a pseudomagnetic field~\cite{Guinea2010a,Vozmediano2010}. Many scanning tunneling microscopy studies in graphene have reported pseudo-Landau levels as signatures of a strain-induced pseudomagnetic field~\cite{Levy2010,Yeh2011,Lu2012,Liu2018,Banerjee2020}, as was also confirmed by means of angle-resolved photoemission~\cite{Nigge2019}. The effects of such an elastic gauge field on the electronic transport properties of graphene are actively being investigated, particularly those related to the control of the valley degree of freedom: graphene valleytronics~\cite{Settnes2016,Stegmann2018,Sandler2020,He2020}. In general, nonuniform strained graphene opens new opportunities to research exotic and in some cases unique behaviors, such as a metal-insulator transition~\cite{Tang2015,Sorella2018}, a fractal spectrum~\cite{Naumis2014,Pedro2015}, superconducting states~\cite{Uchoa2013,Kauppila2016}, and the quantum Hall effect~\cite{Wagner2019,Sela2020}.

In addition to the mentioned strain-induced pseudomagnetic field, another known effect that arises from nonuniform strains is a position-dependent Fermi velocity (PDFV) \cite{FJ2012}, which might induce confinement effects \cite{Oliva2015a,Portnoi2017,Herrmann2018,Alonso2020,Roy2020}. 
Evidence of the PDFV effect in strained graphene has been detected through scanning tunneling spectroscopy (STS) \cite{Hui2013,Jang2014}. These experiments are based on the fact that the slope of V-shaped STS spectra shows variations at different positions of the sample if the Fermi velocity is spatially varying. Alternatively, Landau-level spectroscopy measurements can be used to confirm PDFV effects \cite{Storz2016,Oliva2018a}. Fingerprints of a PDFV on the Landau level spectrum have been theoretically addressed \cite{Oliva2018a,Khaidukov2016,Debus2018}. For example, from tight-binding and Dirac approaches, it was demonstrated in terms of a PDFV that nonuniform uniaxial strains (ripples) produce  position-dependent local density-of-states peaks of graphene under an external uniform magnetic field \cite{Oliva2018a}. Otherwise, Landau levels in rippled graphene are shifted towards lower energies proportionally to the average deformation \cite{Debus2018}.

More recently, Lantagne \emph{et al.} \cite{Franz2020} drew attention to the possibility of achieving spatially separated valley currents and a valley analog to the chiral anomaly in graphene nanoribbons under uniaxial nonuniform strain. Their findings are based on the following remarkable feature: In the presence of certain uniform strain-induced pseudomagnetic fields, the resulting pseudo-Landau levels in graphene are not flat but disperse linearly for small wave-vectors away from the Dirac point, with an opposite slope between the two valleys. As previously reported in the literature \cite{Salerno2015,Salermo2017}, such dispersive behavior of the pseudo-Landau levels was explained by Lantagne and colleagues \cite{Franz2020} as a consequence of a PDFV, confirming that its presence has important physical consequences. However, they do not discuss the possible magnetic effect of a PDFV. The present paper is devoted to demonstrating that, even in the absence of a strain-induced pseudomagnetic field, a PDFV may induce by itself a sort of magnetic field (and therefore a Lorentz-like force) on the charge carriers in strained graphene. 

For the sake of generality, we present this work in the broader context of strained honeycomb lattices, i.e., beyond strained graphene. Nowadays, synthetic systems with honeycomb lattices are artificially created to mimic the behavior of Dirac quasiparticles, such as molecular graphene \cite{Gomes2012}, ultracold atoms \cite{Tarruell2012}, photonic lattices \cite{Rechtsman2013,Rechtsman2020},  polaritonic systems \cite{Klembt2018,Mann2020}, and acoustic structures \cite{Yang2017,Zhang2019}. These artificial graphene-like systems offer the advantage of tuning, in a controlled and independent manner, the hopping parameter between different lattice sites. As a consequence, they open a door to explore the physical effects as well as applications of gauge fields \cite{Gomes2012,Tarruell2012,Rechtsman2020,Rechtsman2013,Klembt2018,Mann2020,Yang2017,Zhang2019}. For instance, for photonic lattices, it has been envisioned that such pseudoelectromagnetic fields will be useful for applications, such as chip-scale nonlinear optics and coupling to quantum emitters, where strong enhancement of light-matter interaction is achieved from the high density-of-states associated with flat bands \cite{Rechtsman2020}. To a large degree, applications of synthetic gauge fields concern phenomena related to the physics of topological quantum matter \cite{Aidelsburger2018,Ozawa2019}. Moreover, it is worth pointing out that such gauge fields also offer the opportunity to investigate novel effects that come from other fields of research, such as high-energy physics \cite{Aidelsburger2018}.

The manuscript is organized as follows. Sec. \ref{M-T} presents the low-energy theory of Dirac fermions on a nonuniform (strained) honeycomb lattice. In Sec. \ref{E-PDFV}, we derive and discuss magnetic effects due to a PDFV on charge carriers.  Further, Sec. \ref{K-T} illustrates the consequences of such PDFV-induced effects on the (Klein) tunneling of electrons through pseudomagnetic barriers. Finally, in Sec.~\ref{Cs}, our conclusions are presented.

\section{Model and theory}\label{M-T}

\begin{figure}[t]
\includegraphics[width=\linewidth]{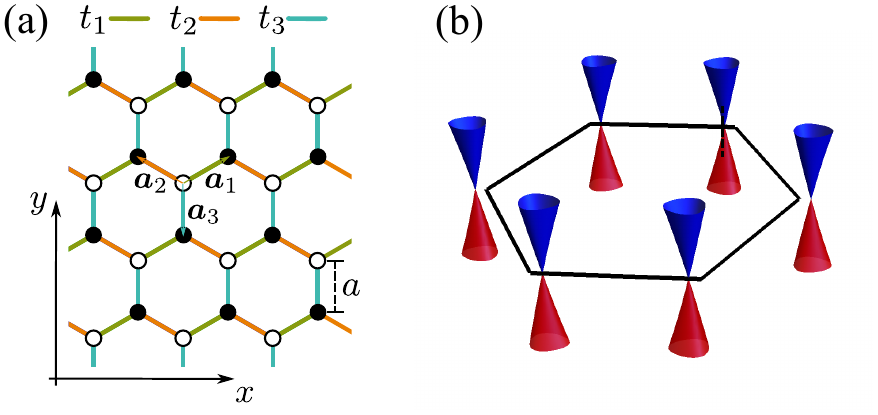}
\caption{(a) Anisotropic honeycomb lattice. Their nearest-neighbor hoppings are dependent on the direction but independent of the position. (b) Illustration of the Dirac cones shift away from the Brillouin zone corners.}\label{fig1}
\end{figure}

\emph{Homogeneous Anisotropy.---}Within a nearest-neighbor tight-binding model for a homogeneous anisotropic honeycomb lattice, e.g. uniformly strained graphene, with  different hoppings $t_{1}$, $t_{2}$ and $t_{3}$ between nearest sites but independent of the position (see Fig.~\ref{fig1}(a)), the Hamiltonian in momentum space can be expressed by a ($2\times2$) matrix of the form \cite{Hasegawa2006}
\begin{equation}\label{TBH}
H(\bm{k})=
-\sum_{n=1}^{3} t_{n}\left(
\begin{array}{cc}
0 &  e^{-\text{i}\bm{k}\cdot\bm{a}_{n}}\\
e^{\text{i}\bm{k}\cdot\bm{a}_{n}} & 0
\end{array}
\right),   
\end{equation}
where $\bm{a}_{n}$ are the nearest-neighbor vectors. This Hamiltonian leads to the dispersion relation of two bands,
\begin{equation}\label{DR}
E(\bm{k})= \pm\left|t_{1}
e^{i\bm{k}\cdot\bm{a}_{1}}+t_{2}
e^{i\bm{k}\cdot\bm{a}_{2}}+t_{3}
e^{i\bm{k}\cdot\bm{a}_{3}}\right|.
\end{equation}

As is well known, for the isotropic (unstrained) case, $t_{1,2,3} = t$, the positions ($\bm{K}^{D}$) of the Dirac cones are located at the corners ($\bm{K}$) of the first Brillouin zone. Nevertheless, for an anisotropic honeycomb lattice, $\bm{K}^{D}$ do not coincide with $\bm{K}$, as illustrated in Fig.~\ref{fig1}(b). In consequence, to obtain the appropriate effective Dirac Hamiltonian, one can no longer expand the Hamiltonian (\ref{TBH}) around $\bm{K}$, but around $\bm{K}^{D}$ \cite{Oliva2013,Oliva2015a,Zubkov2014,Zubkov2015}.

Writing the hoppings as
\begin{equation}
t_{n}=t(1+\delta_{n}), 
\end{equation}
and the nearest-neighbor vectors as
\begin{equation}
\bm{a}_{1}=\frac{a}{2}(\sqrt{3},1), \ \
\bm{a}_{2}=\frac{a}{2}(-\sqrt{3},1),\ \
\bm{a}_{3}=a(0,-1),
\end{equation}
$a$ being the distance between nearest sites, the shift of the Dirac cones (up to first order in the parameters $\delta_{n}$) can be expressed as \cite{Oliva2016}
\begin{equation}\label{KD}
\bm{K}^{D}\approx\bm{K}+\bm{A},
\end{equation}
where 
\begin{equation}\label{VP}
A_{x}=\frac{\tau}{3a}(2\delta_{3}-\delta_{1}-\delta_{2}), \ \
\ \ \ A_{y}=\frac{\tau}{\sqrt{3}a}(\delta_{1}-\delta_{2}),
\end{equation}
and $\tau$ is the valley index of $\bm{K}$ which can take the values $\pm1$ \cite{Bena2009}. Then, by expanding the tight-binding Hamiltonian in momentum space around $\bm{K}^{D}$, with $\bm{k}=\bm{K}^{D} +\bm{q}$, the effective low-energy Hamiltonian of a uniform anisotropic honeycomb lattice becomes \cite{Oliva2016}  
\begin{equation}\label{DH}
\mathcal{H}=\hbar
v_{F}\sum_{i,j}\sigma_{i}(I_{ij}+\Delta_{ij})q_{j},
\end{equation} 
where $v_{F}=3at/2\hbar$, $I_{ij}$ is the $2\times2$ identity matrix, $\bm{\sigma}=(\sigma_{x},\sigma_{y})$ is a vector of Pauli matrices and $\Delta_{ij}$ is the symmetric matrix with the following entries
\begin{equation}\label{Delta}
\Delta_{ij}=
\begin{pmatrix}
\frac{1}{3}(2\delta_{1}+2\delta_{2}-\delta_{3}) & \frac{1}{\sqrt{3}}(\delta_{1}-\delta_{2})\\
\frac{1}{\sqrt{3}}(\delta_{1}-\delta_{2}) & \delta_{3}
\end{pmatrix}.
\end{equation}

Therefore, from  Eq. (\ref{DH}) one can recognize an anisotropic Fermi velocity tensor expressed by,
\begin{equation}\label{GV}
v_{ij} =v_{F}(I_{ij}+\Delta_{ij}), 
\end{equation}
which is directly related to the presence of the deformed Dirac cones with an elliptical cross-section in the relation dispersion due to the hopping anisotropy.

The general expression (\ref{DH}) enables us to obtain straightforwardly the effective Dirac Hamiltonian once the explicit form of the hopping parameters is given. For example, for uniform strained graphene, the hopping parameters $t_{n}$ can be approximated up to first order in the strain tensor $\epsilon_{ij}$ by $t_{n}\approx t[1-(\beta/a^2)\sum_{i,j}a^{i}_{n}\epsilon_{ij}a^{j}_{n}]$, where $t=2.7\,\text{eV}$ and $\beta\sim3$ \cite{Pereira2009,Botello2018}. Then, for graphene under uniform strain, $\delta_{n}=-(\beta/a^2)\sum_{ij}a^{i}_{n}\epsilon_{ij}a^{j}_{n}$, and hence
Eq.~(\ref{Delta}) results in $\Delta_{ij}=-\beta\epsilon_{ij}$. 

\emph{Inhomogeneous Anisotropy.---}Now, let us to consider that the hoppings are functions of the coordinates, i.e. $t_{n}(\bm{r})=t(1+\delta_{n}(\bm{r}))$, varying slowly on the lattice scale. For this general case, the effective low-energy Hamiltonian of the nonuniform anisotropic honeycomb lattice can be obtained from Hamiltonian~(\ref{DH}) as similar to previous considerations for graphene \cite{FJ2012,FJ2013,Zubkov2014,Zubkov2015,Oliva2015a} or Dirac and Weyl semimetals \cite{Z2015,CZ2016} under nonuniform strain. Note that with the simple replacement $\delta_{n}\rightarrow\delta_{n}(\bm{r})$ in the Hamiltonian (\ref{DH}), the terms of the form $\delta_{n}(\bm{r})q_{i}$ break the hermiticity of the resulting Hamiltonian. Therefore, to assure hermiticity, the usual procedure consists of taking Eq.~(\ref{DH}) and doing the symmetric substitution
\begin{equation}
\delta_{n}q_{i}\rightarrow\delta_{n}(\bm{r})\left(-\text{i}\partial_{i}-K^{D}_{i}(\bm{r})\right) - \frac{\text{i}}{2}\partial_{i}\delta_{n}(\bm{r}).
\end{equation}

Using this rule, and considering up to first order in the parameters $\delta_{n}(\bm{r})$, the effective low-energy Hamiltonian of the nonuniform anisotropic honeycomb lattice can be written as
\begin{equation}\label{HG}
\mathcal{H}=-\text{i}\hbar\sum_{i,j}\sigma_{i} v_{ij}\partial_{j} - \hbar
v_{F}\sum_{i}\sigma_{i} A_{i} - \hbar
v_{F}\sum_{i}\sigma_{i} \Gamma_{i},
\end{equation}
where the Fermi velocity tensor $v_{ij}(\bm{r})$ and the gauge field $\bm{A}(\bm{r})$ are respectively given by Eqs.~(\ref{GV}) and (\ref{VP}), but now both are functions of the position vector $\bm{r}=(x,y)$ due to the spatial dependence of the parameters $\delta_{n}(\bm{r})$. The presence of a PDFV is accompanied by the purely imaginary vector field $\bm{\Gamma}(\bm{r})$ given by
\begin{equation}\label{Gamma}
\Gamma_{i}=\frac{\text{i}}{2v_{F}}\sum_{j}\partial_{j}v_{ij}(\bm{r})=\frac{\text{i}}{2}\sum_{j}\partial_{j}\Delta_{ij}(\bm{r}),
\end{equation}
whose specific form guarantees the hermiticity of the Hamiltonian (\ref{HG}). Unlike the well-known gauge field $\bm{A}$, $\bm{\Gamma}$ does not give rise to pseudo Landau levels; however, it has physical significance \cite{FJ2013}.  

If the effect of the hopping variation on the Fermi velocity is not included, Eq.~(\ref{HG}) reduces to the simpler expression $\hbar v_{F}\sum_{i}\sigma_{i}(-\text{i}\partial_{i} - A_i)$. This last Hamiltonian is typically used in the literature to evaluate and/or interpret the strain effects on electronic properties in terms of a pseudomagnetic field perpendicular to the lattice plane and of strength $B_{\text{ps}}=\hbar(\partial_{x}A_{y}-\partial_{y}A_{x})/e$, with $e$ being the quasiparticle charge. 

Moreover, it is important to mention that Hamiltonian (\ref{HG}) resembles the one for the Dirac fermions moving in a curved background. Within a quantum field theory description, the Fermi velocity tensor represents the \emph{vielbein} (tetrad), which encoded the metric of the effective curved space \cite{Zubkov2015,FJ2012,Zubkov2014,Khaidukov2016}. As a consequence, the effects of a PDFV that we discuss in the next section can be interpreted in terms of the emergent vielbein field.
For a detailed correspondence between the tight-binding approach followed in this paper and a quantum field theory approach, see for example Refs. \cite{FJ2012,Zubkov2014}.

\section{Magnetic effects induced by PDFV} \label{E-PDFV}

We now consider two illustrative cases of hopping variation in order to illustrate the corresponding magnetic-like effects due to a PDFV on the quasiparticle dynamics from Eq.~(\ref{HG}).

\subsection{Dispersive pseudo-Landau levels}\label{DLL}

It is well known that certain variations of the hopping parameters lead to a uniform pseudomagnetic field (see Ref.~\cite{Aidelsburger2018} and references therein). For example, this happens for 
\begin{equation}\label{ex2}
\delta_{1}(y)=\delta_{2}(y)=3c_{0}y/4\ \ \text{and}\ \ \delta_{3} = 0,
\end{equation}
since $\bm{A}=(-\tau c_{0}y/2a,0)$, and thus the pseudomagnetic field turns out to be $B_{\text{ps}}=\tau\hbar c_{0}/(2ea)$. In addition to the uniform pseudomagnetic field, the hopping variation (\ref{ex2}) induces also a PDFV tensor given by
\begin{equation}\label{v1}
v_{ij}=v_{F}
\begin{pmatrix}
1 + c_{0}y & 0\\
0 & 1
\end{pmatrix} ,
\end{equation}
such that the Fermi velocity in the $x$-direction depends on the $y$-coordinate. Moreover, note that $\bm{\Gamma}$ is zero. 

\begin{figure}[t]
\includegraphics[width=\linewidth]{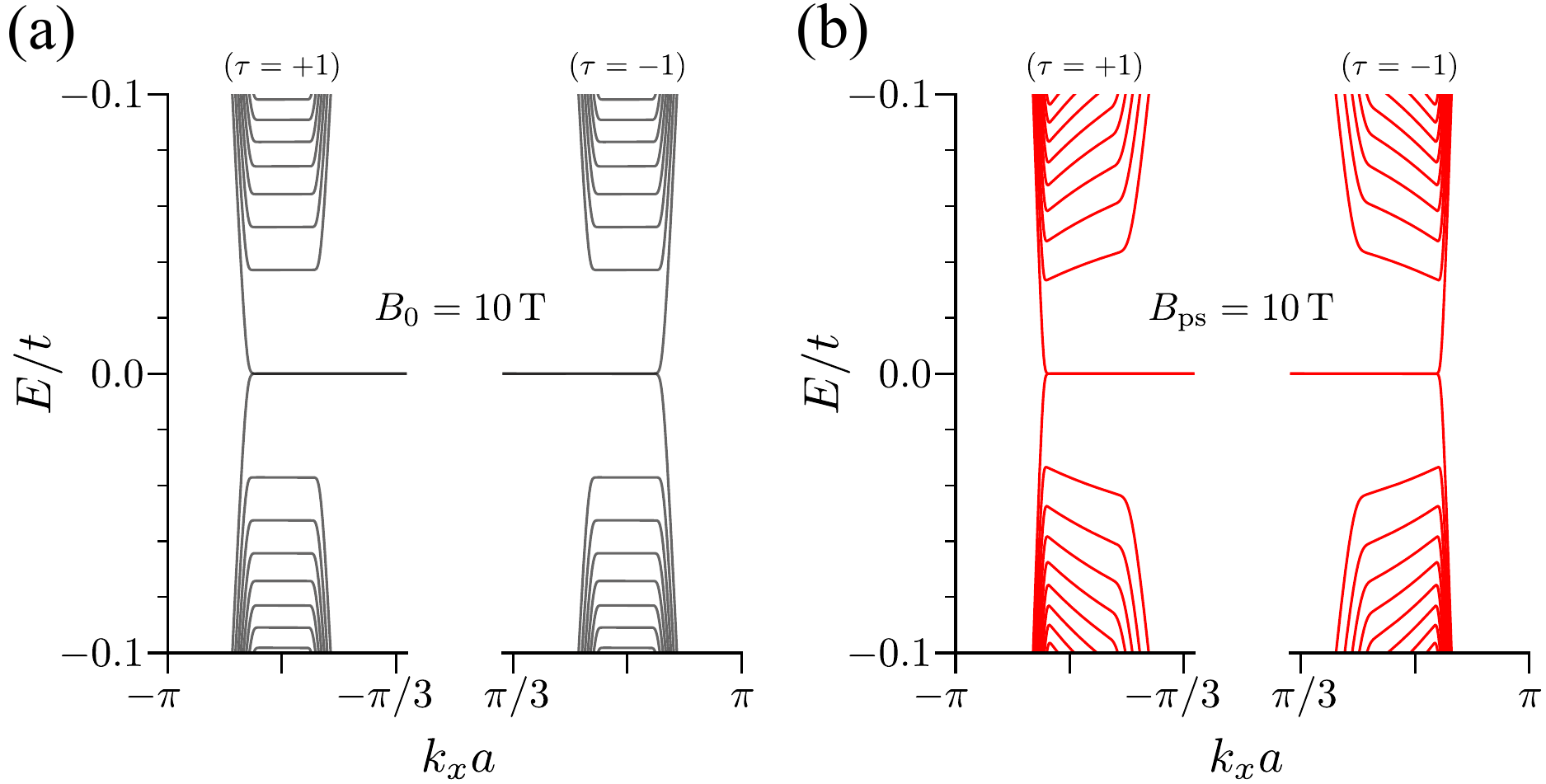}
\caption{(a) Electronic spectrum for a zigzag nanoribbon (of width $\approx 200\,{\rm nm}$) without variation of the hopping parameters and under an external magnetic field of strength $B_0=10\,{\rm T}$. As expected, flat Landau levels near the Dirac points are visible.
(b) Electronic spectrum for a zigzag nanoribbon (of equal width $\approx 200\,{\rm nm}$) in the absence of an external applied magnetic field, but  subject to the variation of the hopping parameters (\ref{ex2}), with a resulting pseudomagnetic field $B_{\rm ps}=10\,{\rm T}$.  Dispersive pseudo-Landau levels with linear dispersion can be recognized. The spectra are obtained from tight-binding calculations.}\label{FigTBH}
\end{figure}

So, for this case, the effective Dirac Hamiltonian (\ref{HG}) reads as 
\begin{equation}
\mathcal{H}=
-\text{i}\hbar v_{F}(1 + c_0 y)\sigma_{x}\partial_{x} -\text{i}\hbar v_{F}\sigma_{y}\partial_{y} + v_{F}e B_{\text{ps}}y\sigma_{x},
\end{equation}
which has translational symmetry along the $x$-direction. As a consequence, the eigenfunctions can be represented as $\Psi(\bm{r})=e^{\text{i}q_{x}x}\Phi(y)$ leading to the following eigenvalues problem 
\begin{eqnarray}
\bigl[(q_{x} +  e\mathcal{B}y/\hbar) - \partial_{y}\bigr]\phi_{2}&=&\frac{E}{\hbar v_{F}}\phi_{1},\nonumber\\
\bigl[(q_{x} +  e\mathcal{B}y/\hbar) + \partial_{y}\bigr]\phi_{1}&=&\frac{E}{\hbar v_{F}}\phi_{2},\label{EP}
\end{eqnarray}
where $\mathcal{B}= B_{\text{ps}} + \hbar q_{x}  c_{0}/e=B_{\text{ps}}(1 + \tau\,2 q_{x}a)$ and $E$ is the energy.
By simple inspection, one can see that Eq.~(\ref{EP}) is analogous to the Landau level problem for massless Dirac fermions in the presence of a uniform magnetic field of strength $\mathcal{B}$, whose spectrum is $E_{n}=\pm\sqrt{2\hbar v_{F}^{2}e\vert\mathcal{B}\vert n}$, with $n$ an integer including zero \cite{McClure1956}. Therefore, the eigenvalues of Eq.~(\ref{EP}) can be expressed as 
\begin{equation}\label{LL}
E_{n,q_{x}}=\pm\hbar v_{F}\sqrt{\vert(c_{0}/a)\vert n}\sqrt{1 + \tau\,2 q_{x}a},
\end{equation}
which can be understood as a result of the combined action of two fields, i.e. the standard pseudomagnetic field $B_{\text{ps}}=\tau\hbar c_{0}/(2ea)$ and an effective magnetic field of strength $\hbar q_{x}  c_{0}/e$, induced by the PDFV (\ref{v1}). Unlike to the standard Landau levels, these resulting levels are no longer flat but dispersive, as coined by Lantagne \emph{et al.} \cite{Franz2020}, because they depend on $q_{x}$ (see Fig.~\ref{FigTBH}). For small $q_{x}$ away from the Dirac point, the dependence is linear ($E_{n,q_{x}}\approx E_n(1+\tau q_{x} a$)) with an opposite slope sign for each valley \cite{Salerno2015,Salermo2017,Franz2020}. As reported in Ref.~\cite{Franz2020}, spatially separated valley currents and a valley analog to the chiral anomaly are a direct consequence of dispersive pseudo-Landau levels near the Dirac points in uniaxially strained graphene nanoribbons.

In Sec. \ref{K-T}, we assess the role that the dispersive pseudo-Landau levels (\ref{LL}) play in the tunneling of electrons through pseudomagnetic barriers.

\subsection{A sublinear dispersion relation}\label{NPF}

Now let us to consider an inhomogeneous honeycomb lattice characterized by the following variation of the hopping parameters
\begin{equation}\label{ex3}
4\delta_{1}(x)=4\delta_{2}(x)=\delta_{3}(x)=c_{0}x,
\end{equation}
whose resulting gauge field is $\bm{A}=(\tau c_{0}x/2a,0)$ and, thus, the corresponding pseudomagnetic field $B_{\text{ps}}$ is zero. Otherwise, the hopping variation (\ref{ex3}) induces a PDFV tensor given by
\begin{equation}\label{fv2}
v_{ij}=v_{F}
\begin{pmatrix}
1 & 0\\
0 & 1 + c_{0}x
\end{pmatrix} ,
\end{equation}
such that the Fermi velocity in the $y$-direction depends on the $x$-coordinate and, besides $\bm{\Gamma}=0$. 

Then, for the considered case (\ref{ex3}), the effective Dirac Hamiltonian (\ref{HG}) can be written as 
\begin{equation}
\mathcal{H}=
-\text{i}\hbar v_{F}\sigma_{x}\partial_{x} -\text{i}\hbar v_{F}(1 + c_0 x)\sigma_{y}\partial_{y},
\end{equation}
with translational symmetry along the $y$-direction. Therefore, the eigenfunctions can be casted as $\Psi(\bm{r})=e^{\text{i}q_{y}y}\Phi(x)$, which leads to the following eigenvalues problem 
\begin{eqnarray}
\bigl[\partial_{x} + (q_{y} + e B_{\text{v}}x/\hbar)\bigr]\phi_{2}&=&\frac{\text{i}E}{\hbar v_{F}}\phi_{1},\nonumber\\
\bigl[\partial_{x} - (q_{y} + e B_{\text{v}}x/\hbar)\bigr]\phi_{1}&=&\frac{\text{i}E}{\hbar v_{F}}\phi_{2},\label{EP3}
\end{eqnarray}
where $e B_{\text{v}}=\hbar q_{y}c_{0}$ is due to the PDFV (\ref{fv2}). As in the previous example, one can recognize that problem (\ref{EP3}) is analogous to the Landau level problem for massless Dirac fermions in the presence of a uniform magnetic field of strength $B_{\text{v}}$. Therefore, the eigenvalues of Eq.~(\ref{EP3}) are given by  
\begin{equation}\label{TLLs}
E_{n,q_{y}}=\pm\hbar v_{F}\sqrt{2\vert q_{y}c_{0}\vert n}.
\end{equation}

In short, in a honeycomb lattice with a PDFV as given by Eq.~(\ref{fv2}), and in the absence of the standard pseudomagnetic field, the dynamics of the Dirac quasiparticles can be understood as if they feel an effective magnetic field of strength $B_{\text{v}}=\hbar q_{y}c_{0}/e$. 
As a consequence, the energy no longer depends linearly on the momentum $q_{y}$, but on the square-root $E\sim\sqrt{\vert q_{y}\vert}$.
 
It is worthwhile to remark that, as mentioned at the end of Sec.~\ref{M-T}, the results here obtained as a consequence of a PDFV could be derived from the equivalent vielbein field. In that case, the spectrum (\ref{TLLs}) would have been called \emph{torsional Landau levels} \cite{Stone2020,Ojanen2020}, since the strength of the vielbein field is referred to as the torsion or torsional magnetic field \cite{Shitade2014,Leigh2014}.

\section{Klein tunneling under pseudomagnetic barriers}\label{K-T}

Because of Klein tunneling, Dirac electrons can tunnel through electrostatic potential barriers without reflection, particularly for normal incidence \cite{Katsnelson2006,Kim09,Stander2009}. This fact makes it difficult to confine Dirac electrons, turning the electronic switching into a big challenge for graphene-based nanoelectronics. In contrast to electrostatic potential barriers, magnetic barriers  (as well as magnetic quantum dots) are able to confine Dirac electrons \cite{DeMartino2007,Ramezani2009,Roy2012,Downing2016}.

\begin{figure*}[t]
\includegraphics[width=0.97\linewidth]{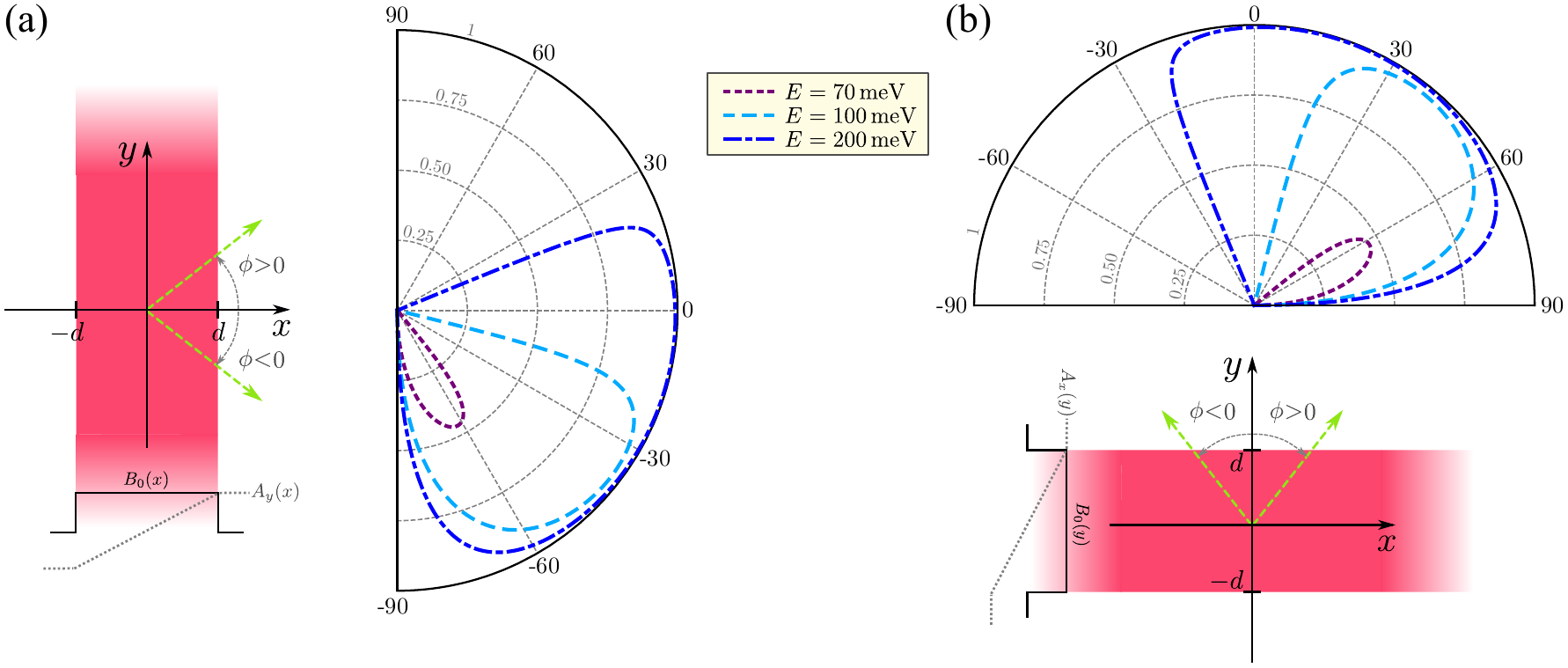}
\caption{(a) On left-side, scattering geometry of Dirac electrons through a magnetic barrier along the $y$-axis with uniform strength $B_{0}$ and width $2d$. On right-side, the respective polar graphs of the transmission probability $T(\phi)$ for $B_{0}=10\,\text{T}$, $d\approx7.1\,\text{nm}$ and different energies. (b) Same as panel (a), but with the magnetic barrier along the $x$-axis.}\label{KT1}
\end{figure*}

Consider a square-well magnetic barrier in the region $\vert x\vert\leq d$, such that the vector potential is given by
\begin{equation}
A_{x}=0\ \ \ \text{and}\ \ \ A_{y}(x)=B_{0}\left\{ \begin{array}{rl}
                                  -d, & \, x < -d \\
                                   x, & \, \vert x\vert\leq d \\
                                   d, & \, x > d \\
                                   \end{array} \right. ,
\end{equation}
with $B_{0}$ being the strength of the external magnetic field within the strip (see Fig.~\ref{KT1}(a)). Proceeding as in Refs.~\cite{DeMartino2007,Ramezani2009}, the electron scattering entering from the left side can be characterized as follows. For $x<-d$, the wave function can be written as \begin{equation}
\Psi_{\text{I}}(x)=
\left(\begin{array}{c} e^{\text{i}q_{x}x} + re^{-\text{i}q_{x}x}\\ e^{\text{i}q_{x}x + \text{i}\phi} -re^{-\text{i}q_{x}x-\text{i}\phi}\end{array}\right), 
\end{equation}
where $\bm{q}=(q_{x},q_{y})$ is the incoming wave vector, $r$ is the reflection amplitude, and $\phi$ is just the kinematic incidence angle $\phi_{i}$, i.e. $\phi=\phi_{i}$, because the gauge-invariant velocity is $\text{\bf{v}}=v_{F}(\cos\phi,\sin\phi)$. Also, note that the wave vector is given by $q_{x}=\varepsilon\cos\phi$ and $q_{y}=\varepsilon\sin\phi + d/l_{B}^{2}$,
with $\varepsilon=E/\hbar v_{F}$ and $l_{B}=\sqrt{\hbar/eB_{0}}$ the magnetic longitude.

Within the region $\vert x\vert\leq d$, the wave function can be expressed as a linear combination of parabolic cylinder functions $D_{\nu}$ (Weber functions), 
\begin{equation}
\Psi_{\text{II}}(x)=
\left(\begin{array}{c}
c_{1}D_{\eta-1} (\text{\small{\textit{X}}}) + c_{2}D_{\eta-1} (-\text{\small{\textit{X}}}) \\
\text{i}\sqrt{2}/(\varepsilon l_{B})[c_{1}D_{\eta} (\text{\small{\textit{X}}}) - c_{2}D_{\eta} (-\text{\small{\textit{X}}})]
\end{array}\right),
\end{equation}
where $c_{1,2}$ are complex constants, $\text{\small{\textit{X}}}=\sqrt{2}(x/l_{B}+q_{y}l_{B})$ and $\eta=(\varepsilon l_{B})^{2}/2$. 

Otherwise, for $x>d$ the transmitted wave function is
\begin{equation}
\Psi_{\text{III}}(x)=t\sqrt{q_{x}/q'_{x}}
\left(\begin{array}{c}
1 \\
e^{\text{i}\phi'} 
\end{array}\right)e^{\text{i}q'_{x}x},
\end{equation}
with $t$ the transmission amplitude and $\phi'$ ($\bm{q}'$) the exit angle (wave vector). From the conservation of $q_{y}$, it results that $\sin\phi'=2d/(\varepsilon l_{B}^2)+\sin\phi$, which implies that for certain incidence angles $\phi$, that fulfill the condition 
\begin{equation}\label{AL1}
-1\leq\sin\phi\leq1-2d/(\varepsilon l_{B}^{2}),
\end{equation}
transmission is possible. However, when $\varepsilon l_{B}\leq d/l_{B}$, any incoming electron is reflected, regardless of the incident angle $\phi$. Then, matching the wave functions and the flux at the interfaces $x=\pm d$, one can obtain in closed form that the transmission amplitude $t$ is given by \cite{DeMartino2007},
\begin{equation}
    t=\frac{2\text{i}\varepsilon l_{B}\sqrt{2q'_{x}/q_{x}}\cos\phi}{e^{\text{i}(q'_{x}+q_{x})d} \mathcal{Q}}(u_2^+ v_2^- + v_2^+ u_2^-),
\end{equation}
where
\begin{align}
 \mathcal{Q}&=(\text{i}\sqrt{2}v_2^+ - \varepsilon l_{B} e^{\text{i}\phi'}u_2^+)
       (-\text{i}\sqrt{2}v_1^- + \varepsilon l_{B} e^{-\text{i}\phi}u_1^-)\nonumber\\
     &\quad +(\text{i}\sqrt{2}v_2^- + \varepsilon l_{B} e^{\text{i}\phi'}u_2^-)
       (\text{i}\sqrt{2}v_1^+ + \varepsilon l_{B} e^{-\text{i}\phi}u_1^+), \nonumber\\
        u^{\pm}_2&=D_{\eta-1}[\pm\sqrt{2}(d/l_{B}+q_{y}l_{B})], \nonumber\\
       v^{\pm}_2&=D_{\eta}[\pm\sqrt{2}(d/l_{B}+q_{y}l_{B})], \nonumber
\end{align}
whereas $u^{\pm}_1$ and $v^{\pm}_1$ follow by letting $d\rightarrow -d$ in the last two expressions, respectively. Finally, the transmission probability is obtained as $T(\phi)=\vert t\vert^2$, which is ultimately a function of the quantities $\phi$, $\varepsilon l_{B}$ and $d/l_{B}$.

Figure~\ref{KT1}(a) shows the $\phi$-dependent transmission  probability $T(\phi)$ of Dirac electrons (in graphene) for different energies through a magnetic barrier with $B_{0}=10\,\text{T}$ and $d=50 a\approx7.1\,\text{nm}$, with $a$ being the carbon-carbon distance. For this barrier, electrons with energies less than $\hbar v_{F}d/l_{B}^{2}\approx62\,\text{meV}$ are reflected, whereas for electrons with energies of $70\,\text{meV}$, $100\,\text{meV}$ and $200\,\text{meV}$, the tunneling is possible for incident angles from $-90\text{\textdegree}$ to the limit angles $-50.6\text{\textdegree}$, $-13.9\text{\textdegree}$ and $22.3\text{\textdegree}$, respectively (see Fig.~\ref{KT1}(a)). If we consider the magnetic field to be $B_{0}=-10\,\text{T}$, the resulting transmission probability can be obtained from the symmetry relation $T(-B_{0},\phi)=T(B_{0},-\phi)$. On the other hand, Fig.~\ref{KT1}(b) depicts the same scenario of electron tunneling (presented in Fig.~\ref{KT1}(a)) but choosing the magnetic barrier along the $x$-axis. 

\subsection{A pseudomagnetic barrier (with dispersive pseudomagnetic field)}

Let us now suppose that the hopping parameters vary as 
\begin{equation}\label{ex2b}
\delta_{1}(y)=\delta_{2}(y)=3c_{0}/4 \left\{ \begin{array}{rl}
                                  -d, & \, y < -d \\
                                   y, & \, \vert y\vert\leq d \\
                                   d, & \, y > d \\
                                   \end{array} \right. ,
\end{equation}
and $\delta_{3} = 0$. In the strip $\vert y\vert < d$, this variation of the hopping parameters induces a standard pseudomagnetic field $B_{\text{ps}}=\tau\hbar c_{0}/(2ea)$ and a Fermi velocity in the $x$-direction depends on the $y$-coordinate. As discussed in Sec.~\ref{DLL}, both effects can be understood as if electrons experience an effective $q_{x}$-dependent magnetic field given by $\mathcal{B}=B_{\text{ps}}(1 + \tau\,2 q_{x}a)$. As a consequence, the previous results about the transmission probability $T$ through a magnetic barrier (along the $x$-axis) can be extended to the present case by replacing $B_{0}$ with $\mathcal{B}$, and thus the magnetic longitude should be redefined as $l_{\mathcal{B}}=\sqrt{\hbar/e\mathcal{B}}$. Following this procedure, $T$ is obtained as a function of the angle $\phi$. Nevertheless, it is important to note that for this case, in the incidence region $y < -d$, the gauge-invariant velocity is $\text{\bf{v}}=v_{F}((1-c_{0}d)\sin\phi,\cos\phi)$. Hence, $\phi$ and the incident angle $\phi_{i}$ no longer coincide. They are related by the expression
\begin{equation}\label{phi1}
 \tan\phi_{i}=(1-c_{0}d)\tan\phi,   
\end{equation}
which can be approximated by $\phi_{i}\approx\phi-\frac{1}{2}c_{0}d\sin2\phi$ (in radians) for $c_{0}d\ll 1$. At last, Eq.~ (\ref{phi1}) allows to parametrically express $T$ as a function of $\phi_{i}$.

\begin{figure}[t]
\includegraphics[width=1\linewidth]{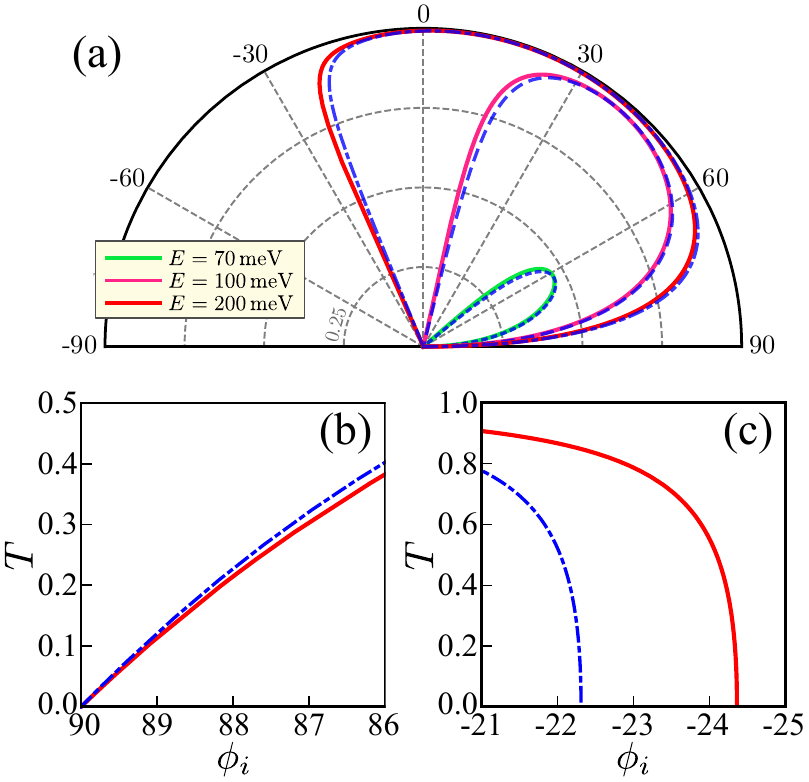}
\caption{(a) Polar graphs of the transmission probability $T(\phi_{i})$
of Dirac electrons with different energies tunneling through a pseudomagnetic barrier along the $x$-axis of width $2d$ and an uniform standard pseudomagnetic field $B_{\text{sp}}=10\,\text{T}$. Solid (dashed blue) lines correspond to the calculation regarding (disregarding) the intrinsic position dependence of the Fermi velocity inside the barrier. Panels (b) and (c) show $T(\phi_{i})$ for different ranges of $\phi_{i}$, but the same energy $E=200\,\text{meV}$. 
}\label{KT2}
\end{figure}

Figure~\ref{KT2}(a) illustrates the transmission probability $T(\phi_{i})$ for electrons (with valley index $\tau=+1$) tunneling through the pseudomagnetic barrier induced by the hopping variation (\ref{ex2b}). We assumed $d=50a$, and the valor of $c_{0}$ was chosen such that $B_{\text{ps}}=10\,\text{T}$. Note that if the position dependence of Fermi velocity is disregarded, and only the standard pseudomagnetic field $B_{\text{ps}}$ is taken into account, then the transmission probability would be as discussed above for a magnetic barrier with $10\,\text{T}$, which is represented by the blue dashed lines in Fig.~\ref{KT2} for reference. Instead, the solid lines in Fig.~\ref{KT2} show the resulting $T(\phi_{i})$ when the position dependence of Fermi velocity is taken into account, i.e., by considering that electrons feel the effective $q_{x}$-dependent magnetic field $\mathcal{B}$. Comparing these results, it can be observed that, in general, the higher the energy of the incident particle through the pseudomagnetic barrier (\ref{ex2b}), the larger is the effect of the PDFV on the transmission probability $T$. In detail, for a given energy (see Fig.~\ref{KT2}(b)), at grazing incidence angles (close to $90\text{\textdegree}$) the resulting $T(\phi_{i})$ is slightly lower because $\mathcal{B}>B_{\text{ps}}$ since $q_{x}>0$ for incidence angles. Otherwise, as illustrated in Fig.~\ref{KT2}(c), at incident angles close to the limit angle of tunneling, $T(\phi_{i})$ turns out to be higher due to $\mathcal{B}<B_{\text{ps}}$, which also leads to a shift of the limit angle itself. In other words, the consideration of the position dependence of Fermi velocity yields a broader spectrum of tunneling angles.

\subsection{A velocity barrier with magnetic features}

Finally, let us consider the variation of the hopping parameters, 
\begin{equation}\label{ex3b}
4\delta_{1}(x)=4\delta_{2}(x)=\delta_{3}(x)=c_{0} \left\{ \begin{array}{rl}
                                  -d, & \, x < -d \\
                                   x, & \, \vert x\vert\leq d \\
                                   d, & \, x > d \\
                                   \end{array} \right. .
\end{equation}

\begin{figure}[t]
\includegraphics[width=0.9\linewidth]{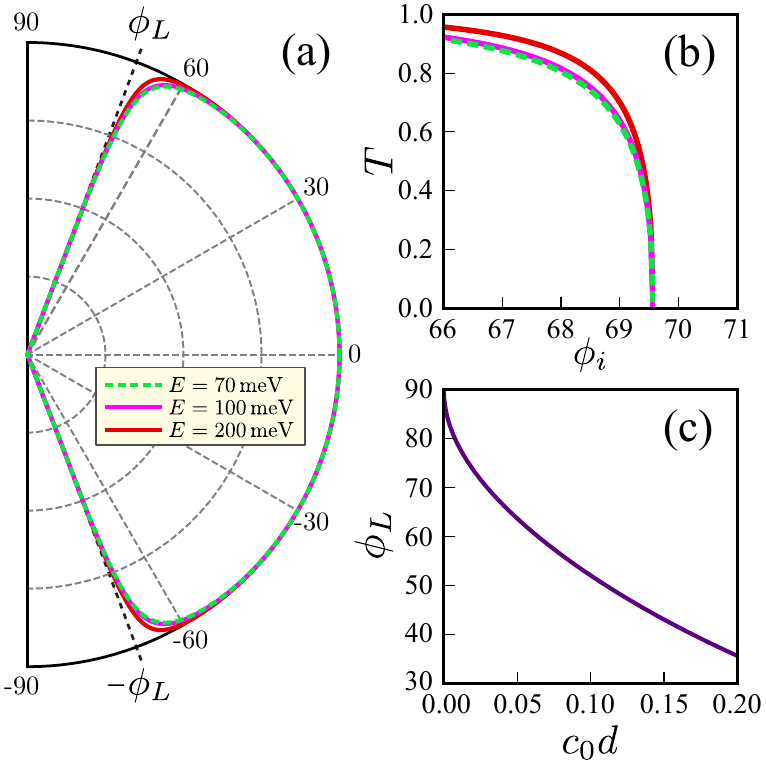}
\caption{(a) Polar graphs of the transmission probability $T(\phi_{i})$ for electrons with different energies tunneling through a magnetic velocity barrier induced by the hopping variation (\ref{ex3b}), such that $c_0 d\approx0.03$. (b) $T(\phi_{i})$ close to the limit incidence angle $\phi_{L}$. (c) Dependence of $\phi_{L}$ as a function of the parameter $c_0 d$. 
}\label{KT3}
\end{figure}

In contrast to the previous example, this does not induce a standard pseudomagnetic field. Note that, as occurred in Sec.~\ref{NPF}, the induced gauge field is of the form $\bm{A}=(A_{x}(x),0)$, so that  $B_{\text{ps}}=\hbar(\partial_{x}A_{y}-\partial_{y}A_{x})/e=0$. However, the hopping variation (\ref{ex3b}) leads to an inhomogeneous Fermi velocity in the $y$-direction, which varies linearly with the $x$-coordinate in the strip $\vert x\vert < d$, from $v_{F}(1-c_0d)$ at $x=-d$ to $v_{F}(1+c_0d)$ at $x=d$. According to Sec.~\ref{NPF}, this last fact mimics the presence of an effective $q_{y}$-dependent magnetic field $B_{\text{v}}=\hbar q_{y} c_{0}/e$ induced by the PDFV. Therefore, regardless of the valley index, an electron that travels from left to right through this kind of magnetic velocity barrier with incidence angle $\phi_{i}>0$ ($\phi_{i}<0$) feels an effective magnetic field $B_{\text{v}}>0$ ($B_{\text{v}}<0$), while for normal incidence $B_{\text{v}}=0$. Therefore, it is expected that the transmission probability fulfills  $T(\phi_{i})=T(-\phi_{i})$ and $T(0)=0$ (see Fig.~\ref{KT3}(a)).

Again the transmission probability $T(\phi)$ can be obtained by using the results for a real magnetic barrier, but now defining the magnetic longitude as $l_{B_{\text{v}}}=\sqrt{\hbar/e\vert B_{\text{v}}\vert}=\sqrt{1/\vert q_{y}c_{0}\vert}$. Also for this case, it is worth noting that $\phi$ and the incident angle $\phi_{i}$ do not coincide because, in the incidence region $x < -d$, the gauge-invariant velocity is $\text{\bf{v}}=v_{F}(\cos\phi,(1-c_{0}d)\sin\phi)$. Hence, both angles are related by expression (\ref{phi1}).

Figure~\ref{KT3}(a) shows $T(\phi_{i})$ for electrons with different energies tunneling through the magnetic velocity barrier produced by the hopping variation (\ref{ex3b}). As in the previous example, here we assumed that $d=50a$ and $c_{0}d\approx0.03$. The reported transmission probability spectrum is notably different from those of electrostatic potential barriers~\cite{Katsnelson2006,Stander2009,Kim09}, magnetic barriers\cite{DeMartino2007,Ramezani2009} and velocity barriers~\cite{Raoux2010,Concha2010}. Its distinguishing feature is that, irrespective of the electron energy, the transmission is almost perfect ($T\approx1$) for the incidence angles in the approximate range $\vert\phi_{i}\vert\lesssim\phi_{L}$, whereas it is completely inhibited ($T=0$) for $\vert\phi_{i}\vert\geq\phi_{L}$ (see Fig.~\ref{KT3}(a)-(b)). The limit angle $\phi_{L}$ is only determined by the barrier parameter $c_0 d$.  By adapting Eq.~(\ref{AL1}) to the present case, we obtain that $\phi_{L}$ is given by
\begin{equation}
 \phi_{L}(c_0d)=\arctan\left[(1-c_0d)\tan \left[ \arcsin\left[\frac{1-c_0d}{1+c_0d}\right] \right] \right], 
\end{equation}
which is plotted (in degrees) in Fig.~\ref{KT3}(c). For instance, $\phi_{L}(0.03)\approx69.8\text{\textdegree}$, $\phi_{L}(0.1)\approx52.0\text{\textdegree}$ and $\phi_{L}(0.2)\approx35.6\text{\textdegree}$. Thus, varying $c_0d$ we could tune the collimation of an electron flux that passes through this magnetic velocity barrier.

\section{Conclusions} \label{Cs}

In closing, we reported new effects due to a PDFV on Dirac fermions in a nonuniform honeycomb lattice, such as, for example, strained graphene or/and  artificial graphene-like systems. It was shown that, for certain spatial dependencies of Fermi velocity, the electrons feel an effective magnetic field. Unlike the known strain-induced pseudomagnetic field, this new sort of magnetic field depends on the electron momentum but not on the valley index. Moreover, we studied fingerprints of a PDFV on the transmission probability of electrons through pseudomagnetic barriers and velocity barriers. For the last ones, we reported anomalous tunneling with magnetic features, which suggests the possibility of using these types of velocity barriers to achieving electron collimation as an alternative to other routes~\cite{Houten88,Louie08,Wang2010}. In general, our results confirmed that the presence of a PDFV can not be obviated because it has important physical consequences, expanding the concept of strain-engineering. Finally, the reported magnetic effects due to a PDFV could be extended to (3D) Dirac and Weyl semimetals, including its artificial versions, where a strain-induced PDFV has also drawn attention~\cite{CZ2016,Zhang2015,Arjona2018}.

\begin{acknowledgments}
This work has been partially supported by CONACyT-Mexico under Grant No. 254414. MOL acknowledges a postdoctoral fellowship from CONACyT--Mexico. JEBV acknowledges funding from PAIP Facultad de Qu\'imica, UNAM (grant 5000-9173).
\end{acknowledgments}

\bibliography{biblio}

\end{document}